\documentclass[aps,twocolumn,pra,showpacs,superscriptaddress,footinbib,tightenlines]{revtex4-2}
\usepackage{amsmath}
\usepackage{graphicx}
\usepackage{color}
\usepackage{xcolor}
\usepackage{amsfonts}
\usepackage[colorlinks,citecolor=blue]{hyperref}
\begin{document}
	\title{Kerr enhanced optomechanical entanglement generation via reservoir design}
	\author{Yan Li}
	\email{These authors contributed equally to this work.}
	\affiliation{Key Laboratory of Low-Dimensional Quantum Structures and Quantum Control of Ministry of Education, Key Laboratory for Matter Microstructure and Function of Hunan Province, Department of Physics and Synergetic Innovation Center for Quantum Effects and Applications, Hunan Normal University, Changsha 410081, China}
	
	\author{Cheng Liu}
	\email{These authors contributed equally to this work.}
	\affiliation{Key Laboratory of Low-Dimensional Quantum Structures and Quantum Control of Ministry of Education, Key Laboratory for Matter Microstructure and Function of Hunan Province, Department of Physics and Synergetic Innovation Center for Quantum Effects and Applications, Hunan Normal University, Changsha 410081, China}
	
	\author{Yu-Hong Liu}
	\email{These authors contributed equally to this work.}
	\affiliation{Key Laboratory of Low-Dimensional Quantum Structures and Quantum Control of Ministry of Education, Key Laboratory for Matter Microstructure and Function of Hunan Province, Department of Physics and Synergetic Innovation Center for Quantum Effects and Applications, Hunan Normal University, Changsha 410081, China}
	
	\author{Yue-Hui Zhou}
	\affiliation{School of Information, Hunan University of Humanities, Science and Technology, Loudi 417000, China}
	
	\author{Jie-Qiao Liao}
	\email{Contact author: jqliao@hunnu.edu.cn}
	\affiliation{Key Laboratory of Low-Dimensional Quantum Structures and Quantum Control of Ministry of Education, Key Laboratory for Matter Microstructure and Function of Hunan Province, Department of Physics and Synergetic Innovation Center for Quantum Effects and Applications, Hunan Normal University, Changsha 410081, China}
	\affiliation{Hunan Research Center of the Basic Discipline for Quantum Effects and Quantum Technologies, Hunan Normal University, Changsha, Hunan 410081, China}
	\affiliation{Institute of Interdisciplinary Studies, Hunan Normal University, Changsha, 410081, China}

\begin{abstract}
Quantum entanglement is a crucial resource in quantum technologies, enabling advancements in quantum computing, quantum communication, and quantum precision measurement. Here, we propose a method to enhance optomechanical entanglement by introducing an optical Kerr nonlinear medium and a squeezed vacuum reservoir of the optomechanical cavity. By performing the displacement and squeezing transformations, the system can be reduced to a standard linearized optomechanical system with normalized driving detuning and linearized-coupling strength, in which the optical and mechanical modes are, respectively, coupled to an optical vacuum bath and a mechanical heat bath. We focus on the entanglement generation in the single stable regime of the system. By evaluating the steady-state logarithm negativity, we find that the optomechanical entanglement can be enhanced within a wide range of the Kerr constant. In addition, the Kerr nonlinearity can extend the stable region, enabling considerable entanglement generation in the blue-sideband parameter region. We also investigate the dependence of the entanglement generation on the average thermal phonon occupation of the mechanical bath and the optical driving amplitude. It is found that the presence of the Kerr nonlinearity allows the generation of optomechanical entanglement even when the thermal phonon occupation of the mechanical bath is as high as 3000. Our findings will provide valuable insights into enhancing fragile quantum resources in quantum systems.
\end{abstract}

\maketitle

\section{Introduction}

Cavity optomechanics primarily investigates the radiation-pressure interaction between optical modes and mechanical modes, along with the novel quantum effects and applications arising from this interaction~\cite{KippenbergSc2008,AspelmeyerPT2012,AspelmeyerRMP2014,bowen2015quantum}. The study of cavity optomechanics holds dual significance in both fundamental theoretical research~\cite{ZurekPT1991,SchwabPT2005} and cutting-edge applied research~\cite{MetcalfeAPR2014}. Not only does it advance the fundamental theoretical exploration of quantum theory, including macroscopic quantum behaviors~\cite{LiaoPRL2016,HongSC2017,FlorianRMP2018} and quantum-classical boundary~\cite{ZurekPT1991,HarocheRMP2013}, but it also advances the applied research value in frontier quantum technologies, especially concerning quantum precision measurement~\cite{CavesRMP1980,VittorioSC2004,LaHayeSC2004} and quantum sensing~\cite{DegenRMP2017,LiuNano2021,BarzanjehNP2022}. \textcolor{black}{In addition, several experiments in cavity optomechanics have demonstrated ultrasensitive detection capabilities near the classical or standard quantum limits~\cite{ForstnerPRL2012,BagciNa2014,MarceloPRX2014}.} \textcolor{black}{In particular, quantum entanglement in a cavity optomechanical system} holds significant research meaning for all these aspects~\cite{VitaliPRL2007,GenesPRA2008,PalomakiSc2013,RiedingerNa2016,MelvynPRL2018,RiedingerNa2018,Ockeloen-KorppiNa2018,Yu2020,JiaoPRL2020,ClarkeNJP2020,NeveuNJP2021,LaiPRL2022,LiuYHPRA2024}, because quantum entanglement is located at the kernel position of quantum theory~\cite{Schcat1935,peres2002quantum}, and it is an important resource in quantum information science~\cite{HorodeckiRMP2009}. \textcolor{black}{Note that  quantum entanglement in other physical systems has also been investigated in recent years~\cite{RaimondRMP2001,LeibfriedRMP2003,LeeSC2011,PanRMP2012,BlaisRMP2021}.}

In general, it is expected to generate strong entanglement for implementing various quantum information processing tasks. In cavity optomechanics, various strategies have been proposed, such as reservoir engineering techniques~\cite{WangPRL2013,ZhangPRA2020,YangPRA2015}, multimode coupling and collective effects~\cite{WangPRA2015,LaiPRA2021,HuangPRA2022}, the injection of squeezed light~\cite{JiaoLPR2024}, and the quantum interference effect~\cite{TianEPRL2013,GenesPRAAM2011}. Usually, quantum entanglement induced by interaction can be enhanced via increasing the coupling strength. For linearized optomechanical interaction, the coupling strength can be enhanced by increasing the driving strength. However, the large driving amplitude is subject to the unstability issue~\cite{GhobadiPRA2011}. This motivates us to explore other methods of enhancing the optomechanical interaction.

In this work, we propose to enhance the linearized optomechanical coupling by introducing a Kerr nonlinearity into the optical cavity. Under the linearization, the Kerr nonlinearity will lead to the two-photon terms $\hat{a}^{\dagger 2}$ and $\hat{a}^2$, which could amplify the optomechanical coupling. To suppress the noise amplified by these terms, we design in advance a squeezed vacuum reservoir to the cavity field. By carefully choosing the squeezing parameters, the additional noise can be completely canceled, reducing the cavity-field environment to a vacuum bath~\cite{XinPRA2015,LemondeNC2016,QinPRL2018,QINPR20241}. Concretely, we apply the displacement and squeezing transformations to realize the linearization and amplification effects~\cite{Zhouyhoe2021}. We also analyze the changes in the scaled detuning and effective coupling strength due to the Kerr nonlinearity. The single- and multi-valued solutions of the cavity-field displacement amplitude are determined to identify the operational regime. The stability is analyzed with the Routh-Hurwitz criterion~\cite{gradshteyn2014table} and we focus on the single-valued parameter range in this work. Note that the Kerr nonlinearity has recently been suggested to enhance the optomechanical effects, such as optical spring~\cite{OtabePRL2024}, optomechanical cooling~\cite{ZoepflPRL2023,diaz2024kerr} and entanglement~\cite{ZhangCPB2013,ChakrabortyJB17}. However, the amplified noise has not been canceled in these studies. In addition, the realization of the Kerr nonlinearity in optomechanical system has been proposed~\cite{GongPRA2009}.

The optomechanical entanglement is evaluated using the logarithmic negativity~\cite{AdessoPRA2004,PlenioPRL2005} based on the covariance matrix. We find that the entanglement behavior, similar to the effective coupling strength, exhibits a non-monotonic dependence on the Kerr constant $ \chi $, namely first increasing and then decreasing with the increase of $ \chi $. In the small $ \chi $ region, the coupling strength reaches its maximum, leading to enhanced optomechanical entanglement. In addition, increasing $ \chi $ could enlarge the steady-state parameter range, allowing the system to remain stable in the blue detuning region, thus enabling entanglement in this region as well. Our work will highlight the potential of the Kerr nonlinearity in controlling optomechanical entanglement, opening new avenues for further investigations into the stability and manipulation of entanglement across various parameter regimes.

The rest of this paper is organized as follows. In Sec.~{\ref{sec2}}, we introduce the physical model and present the Hamitonian, we also introduce the quantum master equation governing the evolution of the open system. In Sec.~{\ref{sec3}}, we perform the linearization of the system and analyze the change of both the normalized detuning and the linearized optomechanical coupling strength induced by the Kerr nonlinearity. In Sec.~{\ref{sec4}}, we study the optomechanical entanglement between the cavity field and the mechanical resonator. Finally, we present some discussions on the experimental implementation of this scheme and conclude this work in Sec.~{\ref{sec5}}.

\section{Physical model}~\label{sec2}
\begin{figure}[tbp]		
	\center\includegraphics[width=0.45\textwidth]{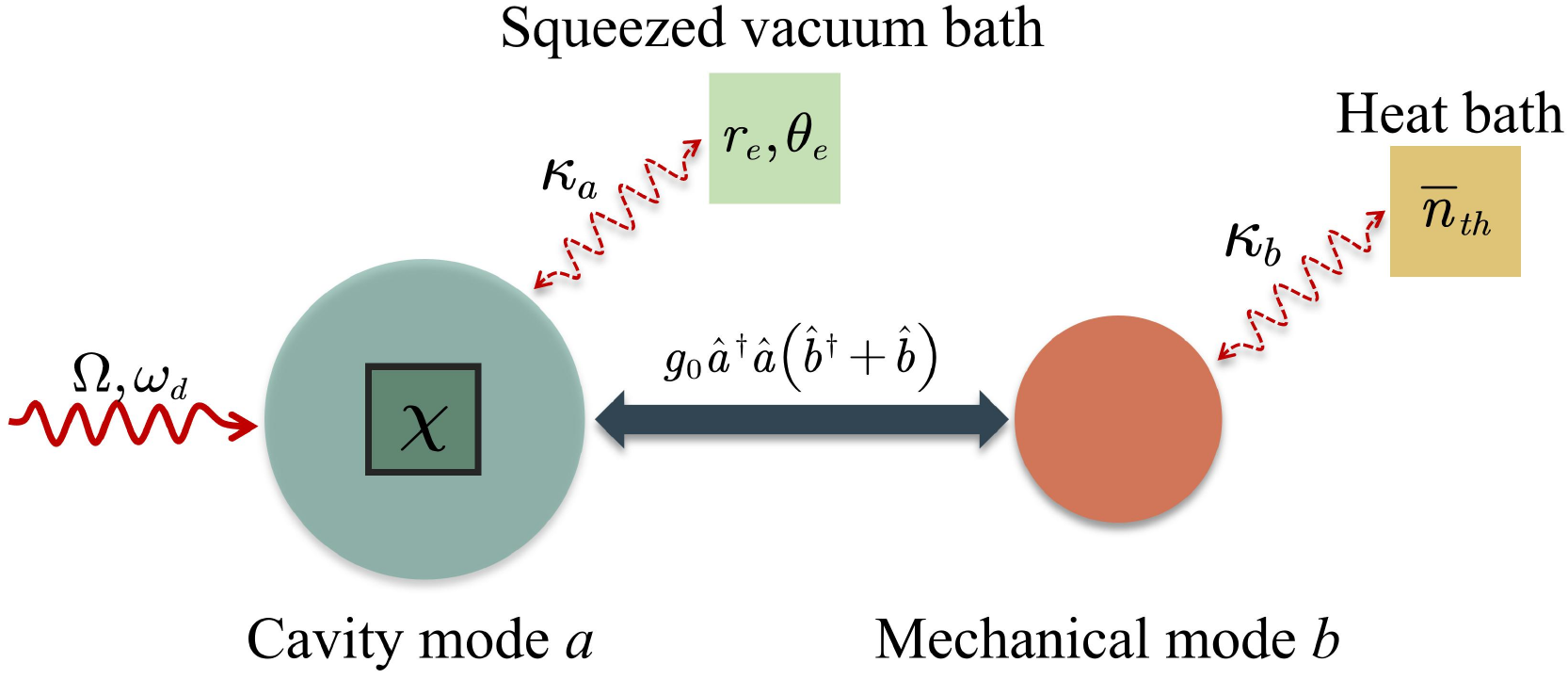}
	\caption{Schematic of the Kerr-cavity optomechanical system consisting of a mechanical mode $b$ optomechanically coupled to a optical mode $a$ containing a Kerr nonlinear medium with nonlinear constant $\chi$. The $g_{0}$ is the single-photon optomechancial coupling strength. The cavity field and the mechanical resonator are, respectively, contacted to a squeezed vacuum reservoir (with decay rate $\kappa_{a}$, squeezing parameter $r_{e}$, and reference phase $\theta_{e}$) and a heat bath (with decay rate $\kappa_{b}$ and average thermal phonon occupation $\bar{n}_{th}$). In addition, the cavity field is driven by a monochromatic field with driving amplitude $\Omega$ and frequency $\omega_d$.}
	\label{Fig1}
\end{figure}
The system under consideration is a Kerr-cavity optomechanical system (Fig.~\ref{Fig1}), which is formed by an optomechanical cavity containing a Kerr nonlinear medium. We consider the adiabatic regime of the cavity field, and hence  focus on a single field mode in the cavity. The cavity field is coupled to the mechanical mode through the radiation-pressure interaction. The cavity is driven by a monochromatic field, and it is connected to a squeezed vacuum bath, while the mechanical mode is coupled to a heat bath. In a rotating frame defined by the unitary transformation operator $\mathrm{exp}(-i\omega_{d}\hat{a}^{\dag}\hat{a}t)$, with $\omega_d$ being the driving frequency, the Hamiltonian of the Kerr cavity optomechanical system reads ($\hbar =1$)
\begin{eqnarray}~\label{eq1}
	\hat{H}_{I}&=&\Delta_{c}\hat{a}^{\dagger }\hat{a}+\omega _{m}\hat{b}^{\dagger }\hat{b}-g_{0}\hat{a}^{\dagger}\hat{a}(\hat{b}^{\dagger }+\hat{b})\notag\nonumber\\
	&&+\chi \hat{a}^{\dagger }\hat{a}\hat{a}^{\dagger }\hat{a}+(\Omega \hat{a}+\Omega^{*}\hat{a}^{\dagger }),
\end{eqnarray}
where $\hat{a}$ $(\hat{a}^{\dagger })$ and $\hat{b}$ $(\hat{b}^{\dagger })$ are, respectively, the annihilation (creation) operators of the cavity-field mode (with resonance frequency $\omega _{c}$) and the mechanical mode (with resonance frequency $\omega _{m}$). The $g_{0}$ term describes the optomechanical coupling, with $g_{0}$ being the single-photon optomechanical coupling strength. The parameter $\chi$ is the Kerr constant, which defines the optical Kerr nonlinearity in the cavity. The $\Omega$ term represents the cavity-field driving with the driving amplitude $\Omega$ and frequency $\omega_d$. The $\Delta _{c}=\omega_{c}-\omega_{d}$ is the detuning of the cavity-field frequency $\omega_c$ with respect to the driving frequency $\omega_d$.

To include the dissipation in the system, we assume that the cavity-field mode is connected to a squeezed vacuum reservoir with the central frequency $\omega_{d}$, and the mechanical mode is coupled to a heat bath. Here, the squeezed vacuum reservoir is introduced to counteract the thermal noise caused by the two-photon terms, which is induced by the Kerr nonlinearity under the linearization. In the Markovian-dissipation regime, the evolution of the system is governed  by the quantum master equation
\begin{eqnarray}~\label{eq2}
	\dot{\hat{\rho}} &=&-i[\hat{H}_{I},\hat{\rho}]+\kappa_{a}(N+1)\hat{\mathcal{D}}[\hat{a}]\hat{\rho } +\kappa_{a}N\hat{\mathcal{D}}[\hat{a}^{\dagger}]\hat{\rho}\notag\\
	&&-\kappa_{a}M\hat{\mathcal{G}}[\hat{a}]\hat{\rho}-\kappa_{a}M^{\ast }\hat{\mathcal{G}}[ \hat{a}^{\dagger}]\hat{\rho}\notag\\
	&&+\kappa_{b}(\bar{n}_{th}+1)\hat{\mathcal{D}}[\hat{b}] \hat{\rho}+\kappa_{b}\bar{n}_{th}\hat{\mathcal{D}}[\hat{b}^{\dagger }]\hat{\rho} ,
\end{eqnarray}
where $\hat{\rho}$ is the density matrix of the optomechanical system and $\hat{H}_{I}$ is given by Eq.~(\ref{eq1}). The $\hat{\mathcal{D}}[\hat{o}]\hat{\rho}=\hat{o}\hat{\rho} \hat{o}^{\dagger }-(\hat{o}^{\dagger}\hat{o}\hat{\rho} +\hat{\rho} \hat{o}^{\dagger }\hat{o})/2$ (for $\hat{o}=\hat{a}, \hat{b}$, $\hat{a}^{\dagger}$, and $\hat{b}^{\dagger}$) and $\hat{\mathcal{G}}[\hat{o}]\hat{\rho}=\hat{o}\hat{\rho} \hat{o}-(\hat{o}\hat{o}\hat{\rho}+\hat{\rho} \hat{o}\hat{o})/2$ are the super-operator acting on the density matrix $\hat{\rho}$. The parameters $\kappa_{a}$ and $\kappa_{b}$ are the decay rates of the cavity-field mode and the mechanical mode, respectively. The $N=\sinh^{2}(r_{e})$ is the mean photon number of the squeezed vacuum reservoir, with $r_{e}$ being the squeezing parameter that characterizes the degree of squeezing. The $M=\cosh(r_{e})\sinh
(r_{e})e^{-i\theta _{e}}$ quantifies the strength of the two-photon correlation, where $\theta_{e}$ is the reference phase of the squeezed field. The $\bar{n}_{th}$ is the thermal excitation number associated with the mechanical mode $b$.

\section{Enhanced optomechanical interaction}~\label{sec3}
In the strong-driving regime of the optomechanical cavity, the dynamics of the system  can be linearized. To this end, we adopt the displacement-transformation method to separate the semiclassical motion and quantum fluctuations~\cite{LiaoJQPRA2020}. Concretely, we make the displacement transformation to quantum master equation~(\ref{eq2}) by introducing the density matrix $\hat{\rho}_d$ in the displaced representation as~\cite{LiaoJQPRA2020,Zhouyhoe2021,LiuPRA2022a}
\begin{eqnarray}~\label{eq3}
	\hat{\rho}_d=\hat{D}_a(\alpha)\hat{D}_b(\beta)\hat{\rho } \hat{D}^{\dagger}_a(\alpha)\hat{D}^{\dagger}_b(\beta),
\end{eqnarray}
where $\hat{D}_a(\alpha)=\exp(\alpha \hat{a}^{\dagger}-\alpha^*{\hat{a}})$ and ${\hat{D}}_b(\beta)=\exp(\beta{\hat{b}}^{\dagger}-\beta^*{\hat{b}})$
are, respectively, the displacement operators of modes $a$ and $b$, with the time-dependent displacement amplitudes $ \alpha(t)$ and $\beta(t)$. In the displaced representation, the quantum master equation (\ref{eq2}) becomes
\begin{align}~\label{eq4}
	\dot{\hat{\rho}}_d =&-i[\hat{H}_{d},\hat{\rho}_d] +\kappa _{a}(N+1)\hat{\mathcal{D}}[\hat{a}]\hat{\rho}_d +\kappa _{a}N\hat{\mathcal{D}}[\hat{a}^{\dagger}] \hat{\rho}_d \notag\nonumber\\
	&-\kappa_{a}M\hat{\mathcal{G}}[\hat{a}] \hat{\rho}_d -\kappa_{a}M^{\ast}\hat{\mathcal{G}}[\hat{a}^{\dagger}] \hat{\rho}_d\notag\nonumber\\
	&+\kappa_{b}(\bar{n}_{th}+1)\hat{\mathcal{D}}[\hat{b}] \hat{\rho}_d		+\kappa_{b}\bar{n}_{th}\hat{\mathcal{D}}[\hat{b}^{\dagger }] \hat{\rho}_d.
\end{align}
In the derivation of Eq.~(\ref{eq4}), we have eliminated the linear term of the operators $\hat{a}$ ($\hat{a}^\dagger$) and $\hat{b}$ ($\hat{b}^\dagger$) by setting their coefficients to be zero. In this way, we obtain the equations of motion of the displacement amplitudes $\alpha(t)$ and $\beta(t)$ as follows:
\begin{subequations}~\label{ban}
	\begin{align}
		\dot{\alpha}&=-\left[\frac{\kappa_a}{2}+i \Delta_c\!+\!i g_0(\beta^*\!+\!\beta)\right] \alpha\!-\!2 i\chi|\alpha|^2 \alpha\!+\!i\Omega, \\
		\dot {\beta} &=-\left(i\omega _{m}+\frac{\kappa_{b}}{2}\right)\beta -ig_{0} \vert \alpha \vert ^{2}.
	\end{align}
\end{subequations}
In this work, we focus on the steady-state properties of the system, i.e., setting $\dot{\alpha}=0$ and $\dot{\beta}=0$, then the steady-state values of the displacement amplitudes $\alpha_{ss}$ and $\beta_{ss}$ can be obtained analytically.

In the steady-state case, the Hamiltonian in Eq.~(\ref{eq4}) can be approximated as
\begin{align}~\label{Hd}
	\hat{H}_{d} =&\Delta_{d}\hat{a}^{\dagger }\hat{a}+\omega _{m}\hat{b}^{\dagger }\hat{b}+\chi\alpha _{ss}^{2}\hat{a}^{\dagger 2}+\chi \alpha_{ss}^{\ast 2}\hat{a}^{2}\nonumber\\	&+(G_{d}\hat{a}^{\dagger}+G_{d}^*\hat{a})(\hat{b}^{\dagger}+\hat{b}).
\end{align}
Here, the normalized detuning $\Delta_{d}$ and linearized opotomechanical coupling strength $G_d$ are defined by
\begin{subequations}~\label{Deltad}
	\begin{align}
		\Delta_{d}=&\Delta_c+4 \chi|\alpha_{s s}|^2+g_0(\beta_{s s}+\beta_{s s}^*), \\
		G_d=&g_{0}\alpha_{ss}.
	\end{align}
\end{subequations}
In order to further eliminate the two-photon terms ($\chi\alpha_{ss}^2\hat{a}^{\dagger2}+ \chi\alpha_{ss}^{*2}\hat{a}^2$) in Hamiltonian~(\ref{Hd}), we introduce the squeezing transformation via~\cite{Zhouyhoe2021}
\begin{equation}
	\hat{\rho}_{sd}=\hat{S}_a^{\dagger}(\xi)\hat{\rho}_d\hat{S}_a(\xi).
\end{equation}
Here, $\hat{\rho}_{sd}$ is the density matrix of the system in the squeezed representation and $\hat{S}_{a}( \xi ) =\exp[(\xi ^{\ast }\hat{a}^{2}-\xi \hat{a}^{\dagger2}) /2]$ is the squeezing operator with $\xi =r\exp(i\varphi)$, where $r$ is the squeezing parameter and $\varphi $ is the reference phase for the squeezed field. In the squeezed representation, the quantum master equation is given by
\begin{align}~\label{dotrho1}
	\dot{\hat{\rho}}_{sd}=&-i[\hat{H}_{sd}, \hat{\rho}_{sd}]\nonumber\\
	&+\kappa_a(N_{s s}+1) \hat{\mathcal{D}}[\hat{a}] {\hat{\rho}}_{sd}+\kappa_a N_{s s} \hat{\mathcal{D}}[\hat{a}^\dagger] {\hat{\rho}}_{sd}\nonumber\\
	&-\kappa_a M_{s s} \hat{\mathcal{G}}[\hat{a}] \hat{\rho}_{sd}-\kappa_a M_{s s}^* \hat{\mathcal{G}}[\hat{a}^\dagger] \hat{\rho}_{sd}\nonumber\\
	&+\kappa_b(\bar{n}_{t h}+1)\hat{\mathcal{D}}[\hat{b}] {\hat{\rho}}_{sd}+\kappa_b \bar{n}_{t h}\hat{\mathcal{D}}[\hat{b}^\dagger] {\hat{\rho}}_{sd},
\end{align}
where the Hamiltonian in the squeezed representation is given by
\begin{align}~\label{Hsd1}
	\hat{H}_{sd}=&\Delta_{sd}\hat{a}^{\dagger }\hat{a}+\omega_{m}\hat{b}^{\dagger}\hat{b}+R\hat{a}^2+R^*\hat{a}^{\dagger2} \nonumber\\
	&+(G_{sd}\hat{a}^{\dagger}+G_{sd}^*\hat{a})(\hat{b}^{\dagger}+\hat{b}).
\end{align}
The parameters $\Delta_{sd}$, $R$, and $G_{sd}$ in Eq.~(\ref{Hsd1}) are defined by
\begin{subequations}~\label{Deltasd}
	\begin{align}
		\Delta_{sd}=&\Delta_d \cosh (2 r)-2 \chi(\alpha_r^2-\alpha_i^2) \sinh (2 r) \cos \varphi\nonumber\\
		&-4 \chi \alpha_r \alpha_i \sinh (2 r) \sin \varphi,\\
		R=&\chi\alpha_{s s}^2 e^{-i \varphi}\sinh ^2 r+\chi\alpha_{s s}^{* 2} e^{i \varphi} \cosh ^2 r-\frac{1}{2} \Delta_d \sinh (2r),
		~\label{pingfang} \\
		G_{sd}=&(G_d\cosh r-G_d^{*} e^{i\varphi} \sinh r)~\label{G1},
	\end{align}
\end{subequations}
where $\alpha_r$ and $\alpha_i$ are, respectively, the real and \textcolor{black}{imaginary} parts of $\alpha_{ss}$, i.e., $\alpha_{ss}=\alpha_r +i\alpha_i$.

It follows from Eq.~(\ref{pingfang}) that the two-photon terms can be eliminated by appropriately choosing the value of $r$ and $\varphi$. Setting $R=0$ leads to
\begin{subequations}~\label{R}
	\begin{align}
		&\!2\chi[(\alpha_r^2-\alpha_i^2) \cos \varphi+2 \alpha_r \alpha_i \sin \varphi]
		\!-\!\Delta_d\tanh(2r)=0,\\
		&\!(\alpha_r^2-\alpha_i^2)\sin\varphi-2\alpha_r \alpha_i\cos\varphi=0.
	\end{align}
\end{subequations}
By solving Eqs.~(\ref{R}), we obtain
\begin{subequations}\label{eq13}	\begin{align}
		\!\!\!\varphi&=\operatorname{acrtan}\left(\frac{2\alpha_r \alpha_i} {\alpha_r^2-\alpha_i^2}\right),\\
		\!\!\!\!\!\!r&=\frac{1} {2}\operatorname{acrtan}\left\{\frac{2\chi}{ \Delta_d}[(\alpha_ {r}^2-\alpha_{i}^2) \operatorname{cos}\varphi+4\alpha_r\alpha_i\operatorname{sin}\varphi]\right\}.
	\end{align}
\end{subequations}
Thus, the two-photon terms ($R\hat{a}^2 + R^* \hat{a}^{\dagger 2}$) in Hamiltonian (\ref{Hsd1}) can be completely eliminated by choosing the values of
$r$ and $\varphi$ given by Eqs.~(\ref{eq13}).

The parameters $N_{ss}$ and $M_{ss}$ in Eq.~(\ref{dotrho1}) are given by
\begin{subequations}~\label{Nss_Mss}
	\begin{align}~\label{Nss_Nss}
		N_{s s}=&\sinh^2 r\cosh^2r_e+\sinh^2r_e\cosh^2r \nonumber\\
		&+\!2\cos(\varphi\!-\!\theta_e)\sinh r\cosh r\sinh r_e\cosh r_e, \\
		M_{ss}=&(\sinh r_e\cosh r\!+e^{-i(\varphi\!-\theta_e)} \sinh r \cosh r_e)\nonumber\\
		&\!\times(\sinh r_e\sinh r\!+e^{i(\varphi\!-\theta_e)} \cosh r_e\cosh r).
	\end{align}
\end{subequations}
We point out that $N_{ss}$ and $M_{ss}$ are the effective parameters associated with the squeezed reservoir of the cavity field, and we can see from Eqs.~(\ref{Nss_Mss}) that the parameters $N_{s s}$ and $M_{s s}$ depend on the parameters $r$, $r_e$, $\varphi$, and $\theta_e$. Here, the parameters $r$ and $\varphi$ have been determined by Eqs.~(\ref{eq13}). Since the squeezed vacuum bath characterized by $r_e$ and $\theta_e$ is introduced to cancel the noise caused by two-photon terms in Eq.~(\ref{Hsd1}), we can choose proper values of $r_e$ and $\theta_e$ such that $N_{ss}=0$ and $M_{ss}=0$.
Then the bath associated with the cavity field is reduced to a vacuum bath, which is the minimal bath experienced by the cavity field.

Below, we analyze how to obtain the solution of $r_e$ and $\theta_e$.
In principle, under the given $r$ and $\varphi$, we can analyze the parameters $N_{s s}$ and $M_{s s}$ as functions of $\theta_e$ and $r_e$. By analyzing Eqs.~(\ref{Nss_Mss}), we find that $N_{ss}=0$ and $M_{ss}=0$ when $\theta_e=\varphi+n\pi$ $(n = 1, 3, 5, \dots)$ and $r_{e}=r$.
It should be emphasized that $N_{s s}=0$ and $M_{s s}=0$ imply that both the thermal noise and the squeezing vacuum noise have been completely suppressed. This working point is very important to observe the optomechanical effects at the single-photon level~\cite{XinPRA2015}. This feature is also the motivation for introducing the squeezing vacuum reservoir, i.e., using a well-designed squeezing vacuum bath to suppress the thermal noise and excitations caused by the two-photon driving.
\begin{figure}[tbp]
	\center\includegraphics[width=0.46\textwidth]{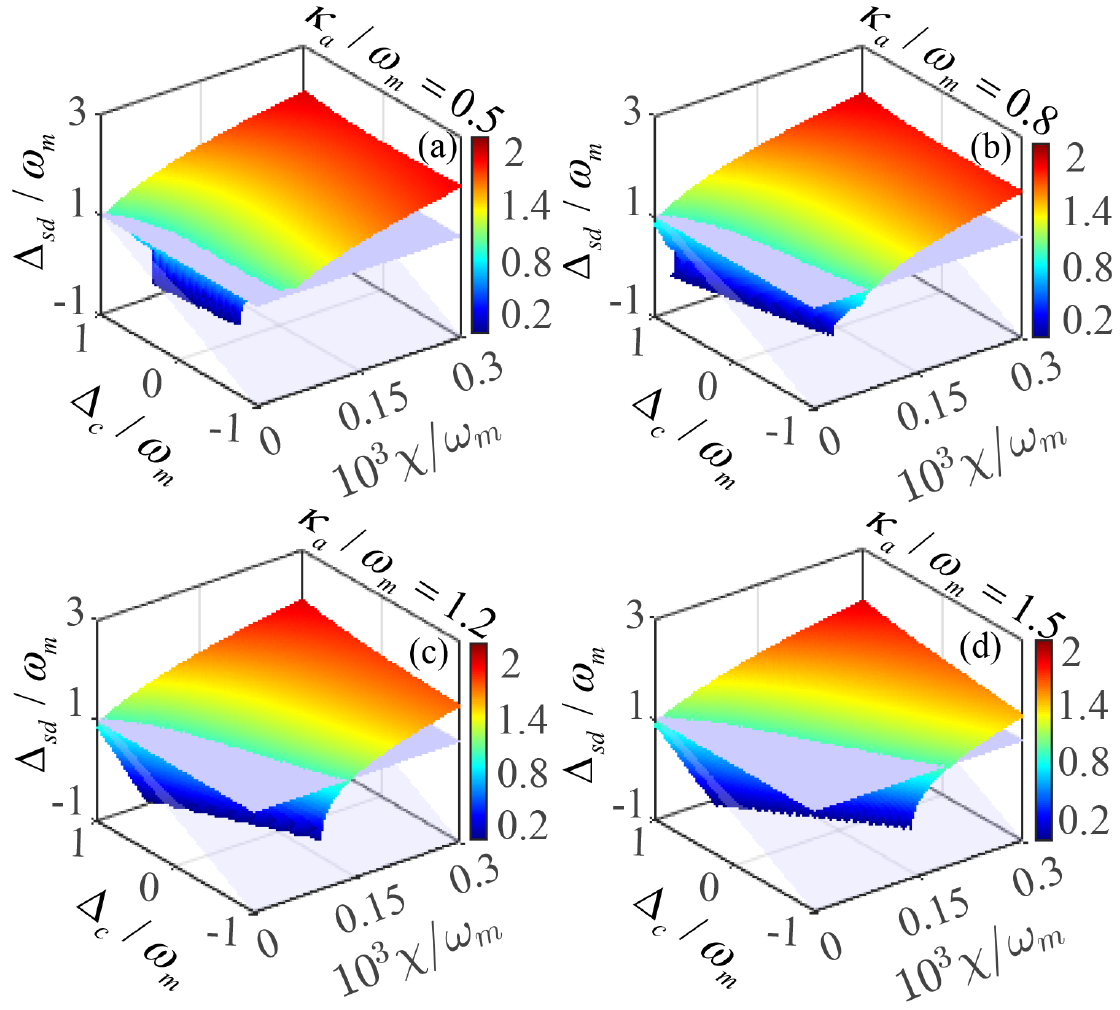}
	\caption{The normalized detuning $\Delta_{sd}/\omega_m$ vs the scaled bare driving detuning $\Delta_{c}/\omega_m$ and the scaled Kerr constant $\chi/\omega_m$ when the cavity-field decay rate $\kappa_{a}$ takes different values: (a) $\kappa_{a}/\omega_m=0.5$, (b) $\kappa_{a}/\omega_m=0.8$, (c) $\kappa_{a}/\omega_m=1.2$, and (d) $\kappa_{a}/\omega_m=1.5$. Other parameters used are $g_{0}/\omega_{m}=0.005$, $\kappa_{b}/\omega_{m}=10^{-5}$, $\Omega/\omega_{m}=50$, and $\bar{n}_{th}=0$. Here, we also show the surfaces corresponding to $\Delta_{sd}/\omega_m=1$ and $\Delta_{sd}=\Delta_{c}$ for comparison.}
	\label{Fig3}
\end{figure}
Under the condition such that $R=0$, $N_{ss}=0$, and $M_{ss}=0$, the quantum master equation (\ref{dotrho1}) is reduced to
\begin{align}~\label{rho11}
	\dot{\hat{\rho}}_{sd}=&-i[\hat{\tilde{H}}_{sd},\hat{\rho}_{sd}]
	+\kappa_a\hat{\mathcal{D}}[\hat{a}]\hat{\rho}_{sd}\nonumber\\ &+\kappa_b(\bar{n}_{th}+1)\hat{\mathcal{D}}[b]\hat{\rho}_{sd} +\kappa_b\bar{n}_{th}\hat{\mathcal{D}}[\hat{b}^\dagger]\hat{\rho}_{sd},
\end{align}
where the Hamiltonian takes the form
\begin{equation}~\label{Hsd}
	\hat{\tilde{H}}_{sd}=\Delta_{sd}\hat{a}^{\dagger}\hat{a}+\omega _{m}\hat{b}^{\dagger}\hat{b}+(G_{sd}\hat{a}^{\dagger}+G_{sd}^*\hat{a})(\hat{b}^{\dagger}+\hat{b}).
\end{equation}
It can be seen from Eq.~(\ref{Hsd}) that, after the squeezing transformation, the system is reduced to an effective and typical linearized optomechanical system. Here, $\Delta_{sd}$ is the effective driving detuning and $G_{sd}$ is the linearized optomechanical coupling strength. In addition, the cavity field is effectively contacted to a vacuum bath, and the mechanical resonator is coupled to a heat bath.

\textcolor{black}{Although the Hamiltonian in Eq.~(\ref{Hsd}) takes the form of a standard linearized optomechanical Hamiltonian, it is important to emphasize that the effective parameters $\Delta_{sd}$ and $G_{sd}$ are nontrivially modified by the Kerr nonlinearity and the squeezed vacuum reservoir. The displacement and squeezing transformations reshape the structure of the parameter space, enabling the system to access a new working point of the larger optomechanical entanglement. Therefore, our method does not simply reproduce a standard linearized optomechanical system but rather provides an experimentally feasible route to expand the entanglement-enhancing parameter regime while ensuring dynamical stability.}

\textcolor{black}{We should point out that, in the present scheme, the entangling power~\cite{ZanardiZ2000} associated with the linearized optomechanical interaction will not be enhanced by introducing the Kerr nonlinearity and the squeezed vacuum bath. This is because the quantum master equation (15) describing the transformed optomechanical system takes the same form as that of a standard linearized optomechanical system. Nevertheless, at certain working points, the entanglement generation can be enhanced because an optimal working point can be achieved owing to the nonlinear mapping of the parameter space. In addition, a further study of the entangling power for the cavity optomechanical model deserves to be conducted because it can give the real upper bound of optomechanical entanglement creation, which will be very important for exploiting quantum technology based on the cavity optomechanical platform.}

\begin{figure}[tbp]
	\center\includegraphics[width=0.48\textwidth]{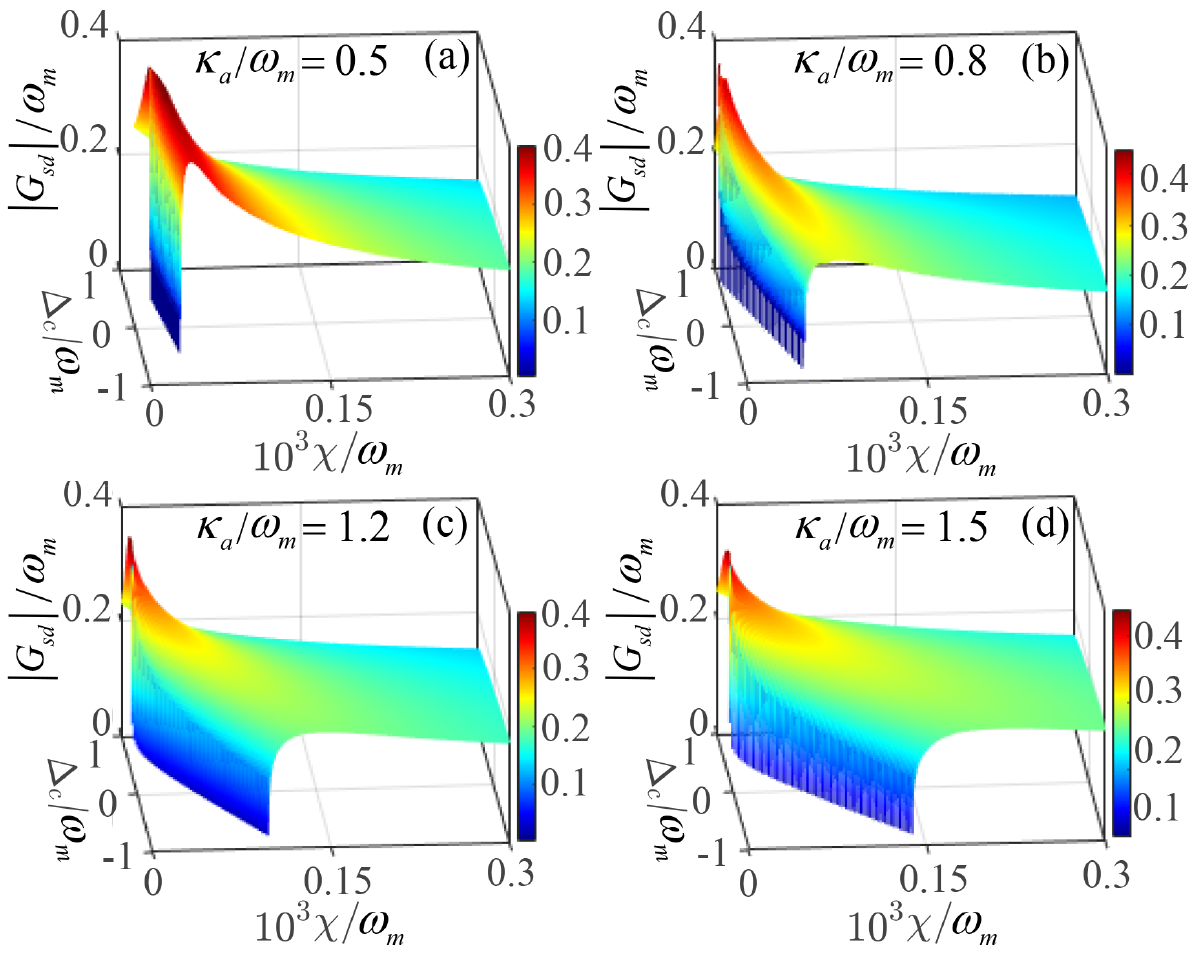}
	\caption{The scaled effective coupling strength $G_{sd}$ vs the scaled bare driving detuning $\Delta_{c}/\omega_m$ and the scaled Kerr constant $\chi/\omega_m$ when the cavity-field decay rate $\kappa_{a}$ takes different values: (a) $\kappa_a/\omega_m=0.5$, (b) $\kappa_a/\omega_m=0.8$, (c) $\kappa_a/\omega_m=1.2$, and (d) $\kappa_a/\omega_m=1.5$. Other parameters are the same as Fig.~\ref{Fig3}.}
	\label{Fig4}
\end{figure}
For generating optomechanical entanglement, both the effective detuning $\Delta_{sd}$ and the effective coupling strength $G_{sd}$ are important parameters. Based on Eqs.~(\ref{ban}), (\ref{Deltad}), (\ref{Deltasd}), and (\ref{R}), we can see that both $\Delta_{sd}$ and $G_{sd}$ depend on the displacement parameters and the squeezing parameters, which are determined by the system parameters.
In Fig.~\ref{Fig3}, we show the dependence of the normalized detuning $\Delta_{sd}/\omega_{m}$ on $\Delta_{c}/\omega_{m}$ and $\chi/\omega_{m}$ for different values of $\kappa_a/\omega_{m}$ in both the sideband resolved and unsolved regimes. Note that Fig.~\ref{Fig3} is plotted only for parameter sets satisfying the steady-state condition, and that the blank area represents the region where the steady-state condition is not met (see Sec.~\ref{sec4}A for details).
We can see from Figs.~\ref{Fig3}(a)-\ref{Fig3}(d) that
the effective driving detuning $\Delta_{sd}/\omega_{m}$ remains positive for these values of $\kappa_a/\omega_{m}$, indicating that the normalized detuning $\Delta_{sd}/\omega_{m}$ caused by both the displacement and squeezing transformations will move towards the red sideband direction. Namely, $\Delta_{sd}/\omega_{m}$ becomes positive within some parameter regions of $\Delta_{c}<0$. For a fixed $\Delta_{c}/\omega_{m}$, the normalized detuning difference $(\Delta_{sd}-\Delta_{c})/\omega_{m}$ increases with the increase of the parameter $\chi$ (as shown by the difference between the surfaces of $\Delta_{sd}/\omega_{m}$ and $\Delta_{c}/\omega_{m}$). In addition, we find that the surface patterns of $\Delta_{sd}/\omega_{m}$ corresponding to the four cases of $\kappa_a/\omega_{m}$ exhibit slight difference. We also find that the red-sideband resonance $\Delta_{sd}/\omega_{m}=1$ corresponds to a relatively small $\chi/\omega_{m}$, and the intersection lines associated with $\Delta_{sd}/\omega_{m}=1$ have similar trend for the four cases of $\kappa_a/\omega_{m}$.

Figures \ref{Fig4}(a)-\ref{Fig4}(d) display the scaled effective coupling strength $G_{sd}/\omega_{m}$ as a function of $\Delta_{c}/\omega_{m}$ and $\chi/\omega_{m}$ for different values of $\kappa_a/\omega_{m}$. Similar to the analysis in Fig.~\ref{Fig3}, these plots are obtained under steady-state conditions. As $\kappa_a/\omega_{m}$ increases, the peak value of $G_{sd}/\omega_{m}$ first increases and then decreases, suggesting a nonmonotonic dependence on $\kappa_a/\omega_{m}$. Increasing $\chi/\omega_{m}$ will broaden the range of $\Delta_{c}/\omega_{m}$ corresponding to a nonzero $G_{sd}/\omega_{m}$, allowing the coupling strength to take significant values over a wider range of detuning. This leads to the possibility for entanglement generation within the blue sideband-parameter region $\Delta_{c}<0$. Additionally, we observe that in both the shallow red- and blue-sideband parameter regions, the coupling strength first increases with $\chi$ and then decreases. In these regions, selecting an appropriate value of $\chi$ can effectively enhance the coupling strength. Notably, the maximum of $G_{sd}$ occurs near the red-sideband resonance $\Delta_{sd}/\omega_{m}=1$ show in Fig.~\ref{Fig3}, emphasizing the importance of near-resonance conditions for achieving strong optomechanical coupling.

\section{Enhanced generation of optomechanical entanglement}~\label{sec4}
In this section, we analyze the stability of the linearized Kerr cavity optomechanical system, derive the steady-state covariance matrix, and evaluate the generated optomechanical entanglement.
\subsection{Analyses of the stability of the system}
For the linearized optomechanical system, its steady state can be characterized by the Gaussian state. Since the equation of motion for the displacement amplitudes $\alpha(t)$ and $\beta(t)$ are nonlinear, rich stability may exist in this system. In our following study of the optomechanical entanglement, we mainly focus on the single stable regime. Therefore, we choose the parameters such that the steady-state displacement amplitudes $\alpha_{ss}$ and $\beta_{ss}$ have a single value. Meanwhile, we use the Routh-Hurwitz criterion to confirm that this steady-state solution is stable. The solution of the amplitudes $\alpha_{ss}$ and $\beta_{ss}$ is obtained by solving Eqs.~(\ref{ban}). To further examine the stability around this working point, we need to evaluate the equation of motion for the first-order moments. To this end, we derive the Heisenberg  equation for $\hat{a}$ and $\hat{b}$ based on the linearized optomechanical Hamiltonian $\hat{\tilde{H}}_{sd}$. To include the dissipations and \textcolor{black}{fluctuations}, we phenomenologically add both the dissipation term and the noise operators into the equations of motion, then we obtain the following Langevin equation,
\begin{subequations}~\label{wen}	
	\begin{align}
		\dot{\hat{a}}=&-\left(i\Delta_{sd}+\frac{\kappa_a}{2}\right)\hat{a}\!-\!iG_{sd} \hat{b}^{\dagger}\!-\!iG_{sd}\hat{b}+\sqrt{\kappa_a}\hat{a}_{in},\\
		\dot{\hat{b}}=&-\left(i\omega_m+\frac{\kappa_b}{2}\right)\hat{b}-iG_{sd}^*\hat{a}-i G_{sd}\hat{a}^{\dagger}+\sqrt{\kappa_b}\hat{b}_{in},
	\end{align}
\end{subequations}
where $\hat{a}_{in}$ and $\hat{b}_{in}$ are, respectively, the noise operators of the modes $a$ and $b$, with the nonzero correlation functions $\langle \hat{a}_{\mathrm{in}}(t) \hat{a}_{\mathrm{in}}^{\dagger}(t^{\prime})\rangle=\delta(t-t^{\prime})$,
$\langle \hat{b}_{in}(t)\hat{b}_{in}^{\dagger}(t^{\prime})\rangle=(\bar{n}_{th}+1) \delta(t-t^{\prime})$, and $\langle \hat{b}_{in}^{\dagger}(t)\hat{b}_{in}(t^{\prime})\rangle=\bar{n}_{th} \delta(t-t^{\prime})$.

\begin{figure}[tbp]
	\center\includegraphics[width=0.47\textwidth]{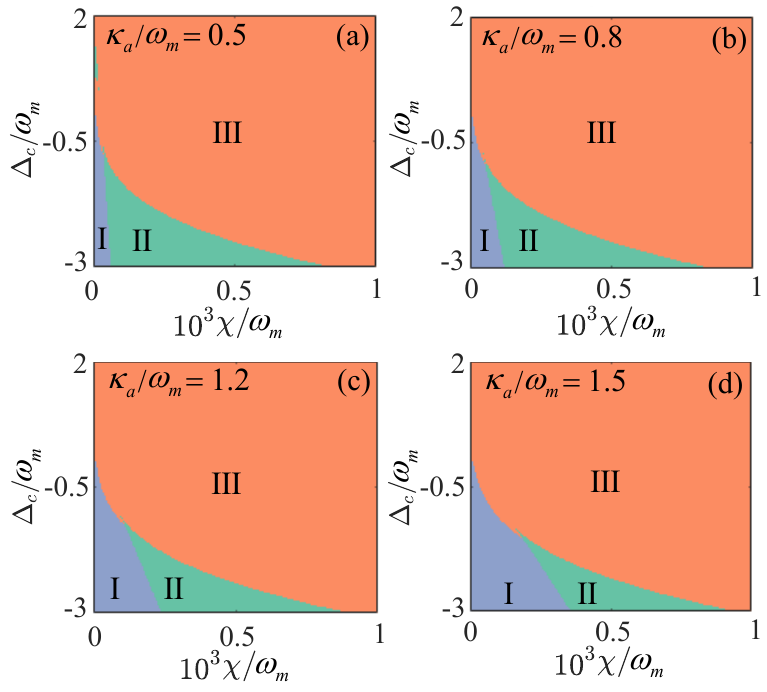}
	\caption{Stability phase diagram of the optomechanical system as a function of $\Delta_c/\omega_{m}$ and $\chi/\omega_{m}$ when (a) $\kappa_{a}/\omega_m=0.5$, (b) $\kappa_{a}/\omega_m=0.8$, (c) $\kappa_{a}/\omega_m=1.2$, and (d) $\kappa_{a}/\omega_m=1.5$. The regions are color-coded as follows: the blue area (region I) corresponds to the single-valued unstable region, the green area (region II) represents the multi-valued region, and the orange area (region III) represents the single-valued stable region. Other parameters are the same as Fig.~\ref{Fig3}.}
	\label{Fig2}
\end{figure}
For convenience, we work in the quadrature representation by introducing the quadrature operators, $\hat{X}_{o}=( \hat{o}^{\dagger }+\hat{o}) /\sqrt{2}$ and $\hat{Y}_{o}=i( \hat{o}^{\dagger}-\hat{o}) /\sqrt{2}$ for $\hat{o}=\hat{a}$ and $\hat{b}$.
In the quadrature representation, we denote $\hat{\mathbf{u}}(t)=( \hat{X}_a,  \hat{Y}_a,  \hat{X}_b,  \hat{Y}_b)^T$ (\textquotedblleft$T$" denotes the matrix transpose); then the Langevin equations can be expressed as
\begin{equation}~\label{u}
	\dot{\hat{\mathbf{u}}} (t)=\mathbf{A}\hat{\mathbf u}(t)+\hat{\mathbf B}(t),
\end{equation}
where the coefficient matrix is introduced as
\begin{eqnarray}\label{eq19}
	\mathbf{A}=\left(\begin{array}{cccc}		
		-\frac{\kappa_a}{2} & \Delta_{sd} & 2 \operatorname{Im} [G_{sd}] & 0 \\
		-\Delta_{sd} & -\frac{\kappa_a}{2} & -2 \operatorname{Re} [G_{sd}] & 0 \\
		0 & 0 & -\frac{\kappa_b}{2} & \omega_m \\
		-2 \operatorname{Re}[G_{sd}] & -2 \operatorname{Im}[G_{sd}] & -\omega_m & -\frac{\kappa_b}{2}
	\end{array}\right),
\end{eqnarray}
and the noise operator vector is defined by
\begin{align}~\label{Nt} {\hat{\mathbf{B}}}(t)=(\hat{X}_{a,in},\hat{Y}_{a,in},\hat{X}_{b,in},\hat{Y}_{b,in})^T.
\end{align}
In Eq.~(\ref{Nt}), the noise operators are defined by $\hat{X}_{o,in}=(\hat{o}_{in}^{\dagger }+\hat{o}_{in}) /\sqrt{2}$, and $\hat{Y}_{o,in}=i(\hat{o}_{in}^{\dagger }-\hat{o}_{in}) /\sqrt{2}$ for $\hat{o}_{in}=\hat{a}_{in}$ and $\hat{b}_{in}$.

The stability of the system can be analyzed by checking the real parts of all the eigenvalues of the coefficient matrix $\mathbf A$. If all the real parts of the eigenvalues are negative, then the system is stable. It can be seen from Eqs.~(\ref{ban}) that the equation could have three solutions under certain parameter conditions, corresponding to three fixed points. If all three fixed points satisfy the steady-state condition, the system enters a tristable regime. If only two fixed points satisfy the steady-state condition, the system is bistable. This implies that the presence of three solutions may lead the system into a multistable regime. Therefore, we exclude the case with three solutions and focus on the single-solution scenario. The first step for clarifying the single stable solution is to distinguish between the multi-solution and single-solution regions. The single-solution cases are then classified as either steady or nonsteady using the Routh-Hurwitz criterion.

In Figs.~\ref{Fig2}(a)-\ref{Fig2}(d), we show the stability phase diagrams of the effective optomechanical Hamiltonian in both the sideband resolved and unsolved regimes. As $\kappa_{a}$ increases, the multi-valued region (green) becomes smaller, and the single-valued unstable region (blue) expands. The single-valued stable region (orange) is only slightly affected, with a slight reduction in the stable area within the deeper blue sideband-parameter region. In addition, as $\chi$ increases, the single-valued stable region expands, reaching into the deep blue sideband-parameter area. In the following sections, we will focus on exploring optomechanical entanglement within the single-valued stable region.

\subsection{The covariance matrix of the system}
For the present linearized optomechanical system, its dynamics can be well described by both the first- and second-order moments. Considering that the first-order moments are initially zero, they will remain zero throughout the subsequent evolution, allowing us to neglect the first-order moments in the following analyses. In this section, we derive the equations of motion of the second-order moments and the steady-state covariance matrix of the system. In terms of the relation
$\partial _{t}\langle \hat{o}_{i}\hat{o}_{j}\rangle
=\text{Tr}(\dot{\hat{\rho}}_{sd}\hat{o}_{i}\hat{o}_{j})$ for $\hat{o}_{i},\hat{o}_{j}$ $\in $ \{$\hat{a}$, $%
\hat{a}^{\dagger }$, $\hat{b}$, $\hat{b}^{\dagger }$\}
and based on the quantum master equation (\ref{rho11}), we can obtain the equation of motion for these second-order moments of the system operators as
\begin{align}
	\dot{\mathbf{X}}(t)=\mathbf{M}(t) \mathbf{X}(t)+\mathbf{N}(t),
\end{align}
where $\mathbf{X}(t)=(\langle \hat{a}^{\dagger} \hat{a}\rangle,\langle \hat{b}^{\dagger} \hat{b}\rangle,\langle \hat{a}^{\dagger} \hat{a}^{\dagger}\rangle,\langle \hat{a} \hat{a}\rangle,\langle \hat{b}^{\dagger} \hat{b}^{\dagger}\rangle,\langle \hat{b} \hat{b}\rangle,\langle \hat{a}^{\dagger}\hat{b}^{\dagger}\rangle,\nonumber\\
\langle \hat{a} \hat{b}\rangle,\langle \hat{a}^{\dagger} \hat{b}\rangle,\langle \hat{a} \hat{b}^{\dagger}\rangle)^{T}$,~$\mathbf{N}(t)=(0, \bar{n}_{th}, 0, 0, 0, 0, iG^*_{sd}, -iG_{sd}, 0, 0)^{T},$ and the coefficient matrix is introduced as $\mathbf{M}(t)=\left(\begin{array}{ll}
	\mathbf{H} & \mathbf{I} \\
	\mathbf{J} & \mathbf{K}
\end{array}\right),$ with
{\small\begin{subequations}~\label{M}
		\begin{align}~\label{H}
			\!\!\mathbf{H}&=\left(\begin{array}{cccc}
				-\kappa_a & 0 & 0 & 0 \\
				0 & -\kappa_b & 0 & 0 \\
				0 & 0 & K_1 & 0 \\
				0 & 0 & 0 & K^*_1
			\end{array}\right),\\
			\!\!\mathbf{I}&=\left(\begin{array}{cccccc}
				0 & 0 & -iG_{sd} & iG^*_{sd} & -iG_{sd} & iG^*_{sd} \\
				0 & 0 & -iG_{sd} & iG^*_{sd} & iG_{sd} & -iG^*_{sd} \\
				0 & 0 & 2iG^*_{sd} & 0 & 2iG^*_{sd} & 0 \\
				0 & 0 & 0 & -2iG_{sd} & 0 & -2iG_{sd}
			\end{array}\right),\\
			\!\!\mathbf{J}&=\left(\begin{array}{cccc}
				0 & 0 & 0 & 0 \\
				0 & 0 & 0 & 0 \\
				iG^*_{sd} & iG^*_{sd} & iG_{sd} & 0 \\
				-iG_{sd} & -iG_{sd} & 0 & -iG^*_{sd} \\
				-iG^*_{sd} & iG^*_{sd} & -iG_{sd} & 0 \\
				iG_{sd} & -iG_{sd} & 0 & iG^*_{sd}
			\end{array}\right),\\
			\!\!\mathbf{K}&=\left(\begin{array}{cccccc}
				K_2 & 0 & 2iG_{sd} & 0 & 0 & 2iG^*_{sd} \\
				0 & K^*_2 & 0 & -2iG^*_{sd} & -2iG_{sd} & 0 \\
				iG^*_{sd} & 0 & K_3 & 0 & 0 & 0\\
				0 & -iG_{sd} & 0 & K^*_3 & 0 & 0 \\
				0 & iG^*_{sd} & 0 & 0 & K_4 & 0 \\
				-iG_{sd} & 0 & 0 & 0 & 0 & K^*_4
			\end{array}\right).
		\end{align}
\end{subequations}}
In Eqs.~(\ref{M}), the variables are introduced as
\begin{subequations}
	\begin{align}
		K_1&=2i\Delta_{sd}-\kappa_a,\\
		K_2&=2i\omega_m-\kappa_b,\\
		K_3&=i\Delta_{sd}+i\omega_m-\frac{\kappa_a+\kappa_b}{2},\\
		K_4&=i\Delta_{sd}-i\omega_m-\frac{\kappa_a+\kappa_b}{2}.
	\end{align}
\end{subequations}

For studying the optomechanical entanglement, it is more convenient to calculate the covariance matrix in the quadrature representation. To this end, we introduce the covariance matrix $\mathbf{V}$ defined by the matrix elements
\begin{align}\label{Vij}
	\mathbf{V_{i,j}}=&\frac{1}{2}[\langle \hat{\mathbf{u}}_{i}(t) \hat{\mathbf{u}}_{j}(t)\rangle+\langle \hat{\mathbf{u}}_{j}(t) \hat{\mathbf{u}}_{i}(t)\rangle]-\langle \hat{\mathbf{u}}_{i}(t)\rangle\langle \hat{\mathbf{u}}_{j}(t)\rangle,
\end{align}
where $\mathbf{u}_i$ for $i=$1-4 are the element of the operator vector $\mathbf{u}$ defined in Eq.~(\ref{u}). The elements of the covariance matrix can be further expressed as a function of the second-order moments~\cite{LiuPRA2024}. In the steady state of the system, the covariance matrix takes the form
\begin{eqnarray}~\label{V}
	\mathbf{V}=\left(\begin{array}{cc}
		\mathbf{V_{O}} & \mathbf{C} \\
		\mathbf{C^T} & \mathbf{V_{M}}
	\end{array}\right),
\end{eqnarray}
where the matrix $\mathbf{V_{O}}$ is associated with the optical mode, $\mathbf{V_{M}}$ is related to the mechanical resonator, and $\mathbf{C}$ characterizes their interaction, which quantifies the bipartite entanglement between the cavity field and the mechanical mode. To quantity the quantum entanglement for the Gaussian steady state of the cavity field and mechanical mode, we employ the logarithm negativity, which is defined by~\cite{AdessoPRA2004,PlenioPRL2005}
\begin{eqnarray}~\label{EN}
	E_{N}=\max [0,-\ln (2\eta^{-})].
\end{eqnarray}
Here, the variable $\eta^{-}$ is defined by
\begin{eqnarray}~\label{eta}
	\eta^{-}=\frac{1}{\sqrt{2}}[\Sigma(\mathbf{V})-\sqrt{\Sigma(\mathbf{V})^2-4 \operatorname{det} \mathbf{V}}]^{1 / 2},
\end{eqnarray}
where $\Sigma(\mathbf{V})=\operatorname{det} (\mathbf{V_{O}})+\operatorname{det}(\mathbf{V_{M}})-2 \operatorname{det}(\mathbf{C}$). Based on the steady-state covariance matrix $\mathbf{V}$, the logarithm negativity can be calculated, and the dependence of the optomechanical entanglement on the system parameters can be analyzed in detail.

\begin{figure}[tbp]	
	\center\includegraphics[width=0.48\textwidth]{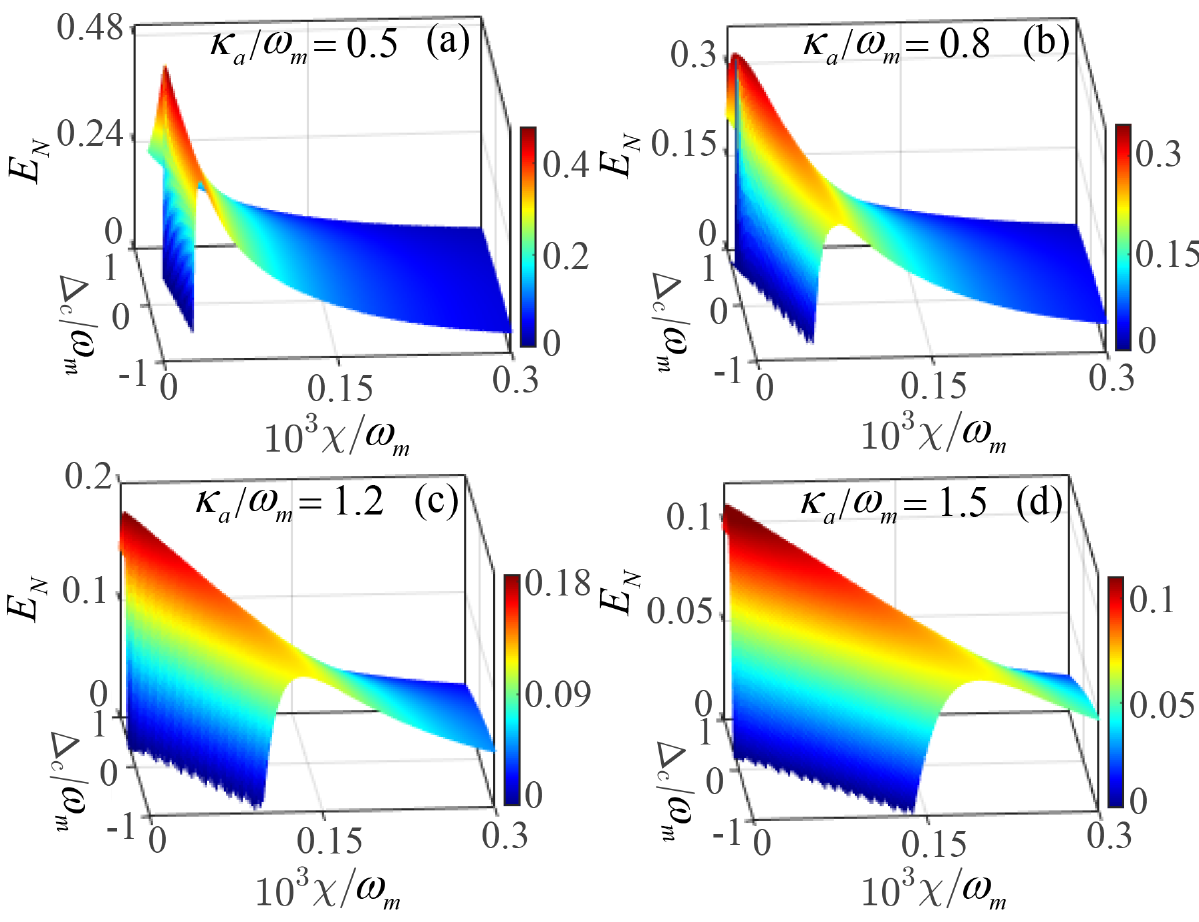}
	\caption{The logarithmic negativity $E_N$
		as a function of $\Delta_{c}/\omega_m$ and $\chi/\omega_m$ when (a) $\kappa_a/\omega_m=0.5$, (b) $\kappa_a/\omega_m=0.8$, (c) $\kappa_a/\omega_m=1.2$, and (d) $\kappa_a/\omega_m=1.5$. Other parameters are the same as those in Fig.~\ref{Fig3}.}
	\label{Fig5}
\end{figure}
\subsection{Enhanced optomechanical entanglement}
In this section, we study the enhanced optomechanical entanglement in the linearized Kerr-cavity optomechanical system. Concretely, we analyze the influence of the system parameters on the entanglement generation. In Figs.~\ref{Fig5}(a)-\ref{Fig5}(d), we show the dependence of the logarithmic negativity on $\Delta_c$ and $\chi$ under different cavity-field decay rates $\kappa_a$. We can see that the peak value of the logarithm negativity is smaller for a larger decay rate $\kappa_a$, and that the optomechanical entanglement remains even in the sideband unresolved regime. Notably, as $\chi$ increases, the large blue sideband-parameter region also becomes suitable for preparing optomechanical entanglement. Moreover, the entanglement exhibits a similar trend as the effective coupling strength $G_{sd}$. By comparing Fig.~\ref{Fig4} with Fig.~\ref{Fig5}, we find that the location in the parameter space spanned over $\Delta_c/\omega_{m}$ and $\chi/\omega_{m}$ corresponding to the peak entanglement is consistent with that for the coupling strength. In particular, in the shallow red- and blue-sideband parameter regions, the $E_{N}$ first increases and then decreases with $\chi$ for a given $\Delta_c$, indicating an enhancement of the entanglement generation induced by the Kerr nonlinearity. This is because the entanglement is determined by both the effective driving detuning $\Delta_{sd}$ and the effective coupling strength $G_{sd}$. As $\chi$ increases, the rapid decline in entanglement is mainly due to the deviation of the detuning $\Delta_{sd}$ from the near-resonant region, causing a faster decrease in entanglement compared to the reduction in effective coupling strength.

\begin{figure}[tbp]
	\center\includegraphics[width=0.48\textwidth]{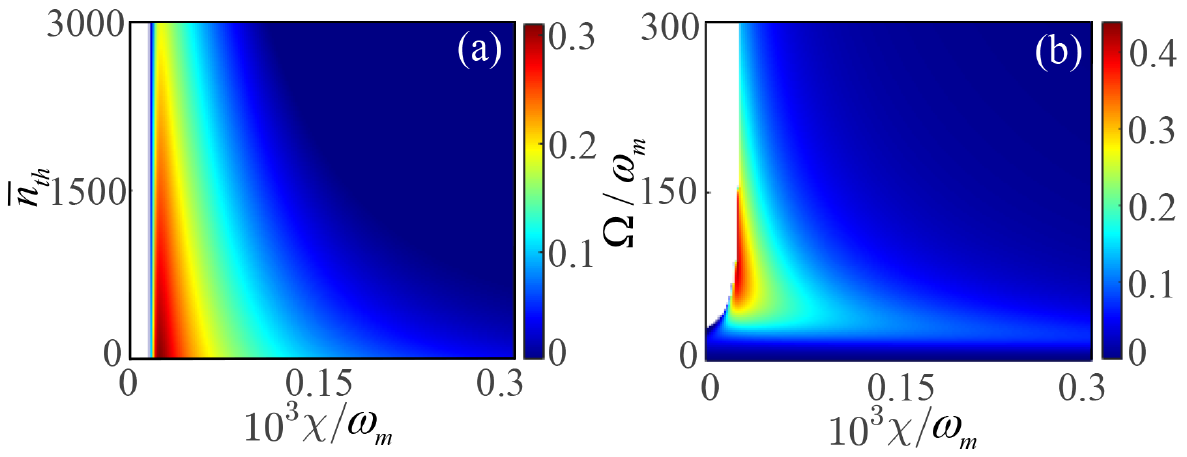}
	\caption{(a) The logarithmic negativity $E_N$ as a function of the average thermal phonon occupation $\bar{n}_{th}$ and the scaled Kerr constant $\chi/\omega_m$. (b) The logarithmic negativity $E_N$ as a function of the driving amplitude $\Omega$ and $\chi/\omega_m$. The parameters used are $\kappa_{a}/\omega_{m}=0.8$ and $\Delta_{c}/\omega_{m}=0.3$. Other parameters are the same as Fig.~\ref{Fig3}.}
	\label{Fig6}
\end{figure}

We also investigate the influence of both the environmental thermal phonon occupation and optical driving amplitude on the entanglement generation performance. Figure~\ref{Fig6}(a) shows the  logarithmic negativity versus the average thermal phonon occupation $\bar{n}_{th}$ and the Kerr constant $\chi$. It can be seen that, as $\chi$ increases, the entanglement initially increases and then decreases. For some values of $\chi$, even when the average thermal phonon occupation $\bar{n}_{th}$ reaches 3000, the steady-state optomechanical entanglement still remains. We point out that athough the $n_{th}$ is very large, the average phonon number in the mechanical mode is small. This is because the working point of the system is very useful for the sideband cooling, and hence the ground-state cooling of the mechanical resonator can be realized. We checked the average phonon number in the mechanical mode and find that $\langle b^{\dagger}b\rangle_{ss}\approx0.5$. We also find that, for a given $n_{th}$, the $E_{N}$ first increases and then decreases with the increase  of $\chi$. This implies that the introduction of $\chi$ improves the ability of the system to resist thermal noise. This phenomenon can be explained based on the dependence of $G_{sd}$ on the Kerr constant $\chi$. Compared to the case of $\chi=0$, the coupling strength $G_{sd}/\omega_{m}$ is largely enhanced for a wide range of  $\Delta_{c}/\omega_{m}$. In Fig.~\ref{Fig6}(b), we show the logarithmic negativity versus the optical driving amplitude $\Omega$ and Kerr constant $\chi$.  It can be observed that as the driving amplitude increases, the entanglement first increases and then decreases, suggesting that stronger driving does not always lead to greater entanglement. Similarly, the peak value of $E_{N}$ appears around  $\chi/\omega_{m}\approx2.4\times10^{-5}$.

\begin{figure}[tbp]
	\center\includegraphics[width=0.48\textwidth]{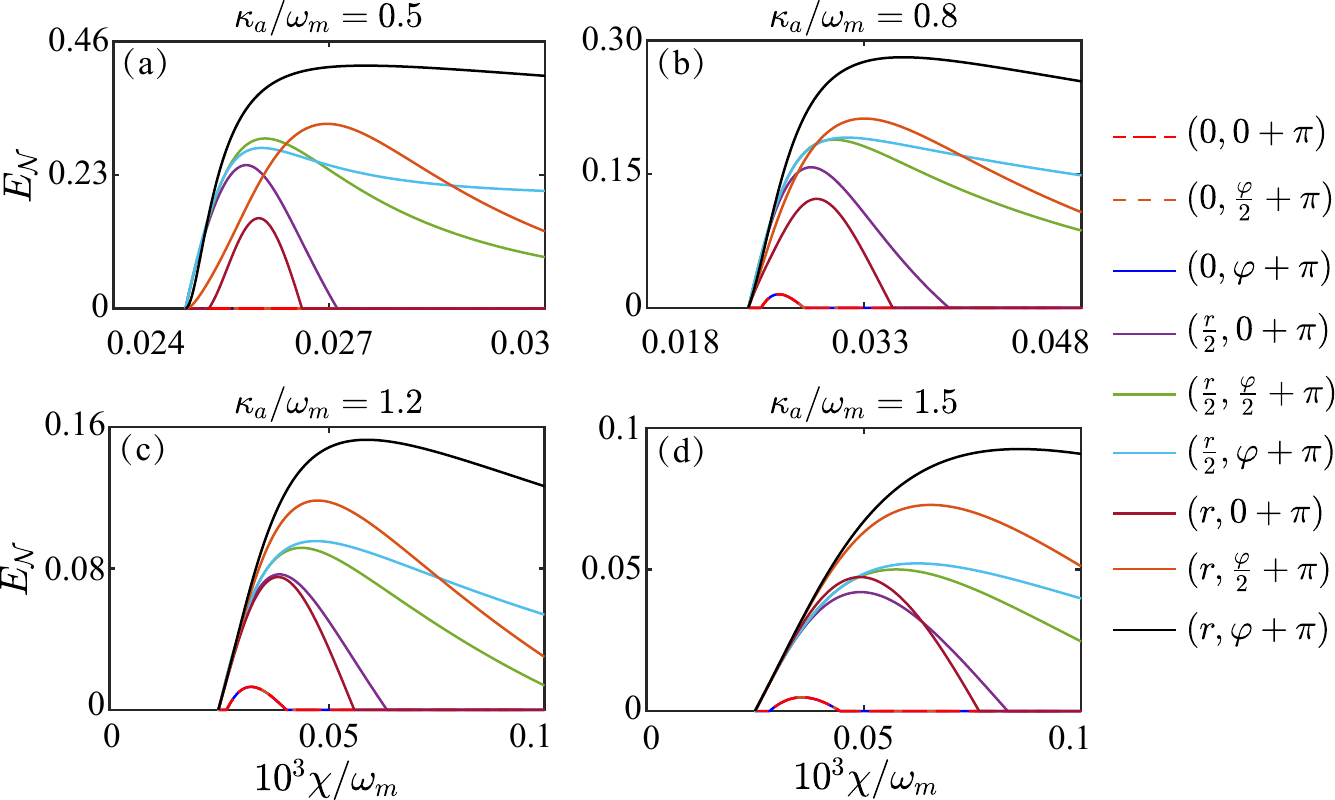}
	\caption{Logarithmic negativity $E_{\mathcal{N}}$ as a function of $\chi/\omega_{m}$
		for different values of the parameters $(r_{e},\theta_{e})$,
		under the four cavity damping rates: (a) $\kappa_{a}/\omega_{m}=0.5$,
		(b) $\kappa_{a}/\omega_{m}=0.8$, (c) $\kappa_{a}/\omega_{m}=1.2$,
		and (d) $\kappa_{a}/\omega_{m}=1.5$. Each panel includes nine curves
		for $r_{e}=\{0, r/2, r\}$ and $\theta_{e}=\{0+\pi,\varphi/2+\pi,\varphi+\pi\}$,
		illustrating the impact of the squeezed vacuum reservoir on the steady-state
		optomechanical entanglement. Here,
		 $\Delta_{c}/\omega_{m}=0$, and other parameters are the same as Fig.~\ref{Fig3}.}
	\label{Fig7}
\end{figure}
	
\textcolor{black}{In the above analyses, we demonstrate the enhanced optomechanical entanglement based on Eq.~(\ref{Hsd}), where the two-photon noise terms induced by the Kerr interaction are completely canceled by the introduced squeezed reservoir [the parameters $N_{ss}$ and $M_{ss}$ in Eqs.~(\ref{Nss_Mss}) are equal to zero]. Next, we investigate the optomechanical entanglement for the case where the amplitude or phase of the squeezed reservoir does not give rise to perfect cancellation of these two-photon noise terms. To this end, we numerically calculate the logarithm negativity to characterize the optomechanical entanglement for various amplitudes $r_{e}$ and phases $\theta_{e}$ under these four cavity damping rates $\kappa_{a}/\omega_{m}=0.5, 0.8, 1.2$, and $1.5$, as shown in Fig.~\ref{Fig7}.} 

\textcolor{black}{In Fig.~\ref{Fig7}, we can see that the logarithm negativity reaches its maximum for the parameter $(r_{e},\theta_{e})=(r,\varphi+\pi)$, while it is very small for the parameter $(r_{e},\theta_{e})=(0,0+\pi)$. The results mean that the optomechanical entanglement is optimized when the two-photon term induced by the Kerr interaction is completely canceled by the squeezed vacuum reservoir, while the absence of such a reservoir leads to a strong suppression for the optomechanical entanglement. Meanwhile, the intermediate configurations such as $(r_{e},\theta_{e})=(0,\varphi/2+\pi),(0,\varphi+\pi),(r/2,0+\pi),(r/2,\varphi/2+\pi),(r/2,\varphi+\pi),(r,0+\pi),$ and $(r,\varphi/2+\pi)$ clearly show reduced logarithm negativity, which indicate that, though the optomechanical entanglement is sensitive to mismatches in the squeezing parameters $r_{e}$ and $\theta_{e}$, the system still tolerates moderate deviations. This highlights the experimental feasibility of the proposed scheme, even with the imperfect squeezed vacuum reservoir.}

\section{Discussions and Conclusion}~\label{sec5}
Finally, we present some discussions on the experimental implementation of this scheme. In our scheme, there are four elements: the optomechanical coupling between the field and the mechanical motion, the Kerr interaction of the cavity field, the designed squeezed vacuum bath of the cavity field, and the monochromatic driving of the cavity field. To implement the scheme, the physical system candidate should be able to realize these four elements. The optomechanical coupling can now be implemented in many physical systems, including optical
microresonators~\cite{GroblacherNP2009,ParkNp2009,SchliesserNp2009,VerhagenNa2012}, electromechanical systems~\cite{DobrindtPRL2008,HertzbergNp2010,MasselNa2011,TeufelNa2011,MasselNc2012}, photonic crystal nanobeams~\cite{EichenfieldNa2009,EichenfieldJasperNa2009}, and Fabry-Pérot cavities~\cite{GroblacherNa2009}.
The monochromatic driving for the cavity field is experimentally accessible in many physical systems. Therefore, the key elements are the Kerr interaction and the squeezed vacuum bath. Usually, the Kerr interaction can be induced by coupling the cavity field to an $N$-type four-level atomic system. It has been reported that the Kerr constant can reach $\chi/2\pi\approx 0.16$~MHz~\cite{ZoepflPRL2023}.
In addition, we point out that the Kerr interaction of the cavity field exists in the circuit-QED system~\cite{KirchmairNa2013} and the Kerr interaction has been realized in cavity magnomechanical systems~\cite{HongSC2016,WangPRB2016}. In particular, several experiments have reported the realization of the Kerr interaction~\cite{MasselPRL2014,EichlerPRL2014,ArrangoizPRX2018}. \textcolor{black}{Notably, recent experiments have demonstrated Kerr-enhanced optomechanical interactions in both microwave and optical systems. In particular, superconducting circuit electromechanics~\cite{BothnerComPhys2022} and optical Fabry-Pérot cavities at 1064~nm~\cite{OtabePRL2024} both show the compatibility of the Kerr effect with optomechanical platforms.}
For the squeezed vacuum bath, it can be created by injecting a squeezed vacuum field into the cavity mode. Based on the above analyses, our scheme should be within the reach of current and near-future experimental conditions.

In conclusion, we have proposed a scheme to enhance the optomechanical entanglement. This is achieved by introducing both the Kerr interaction in the cavity field and the squeezed vacuum bath. Under the strong driving of the cavity field, the two-photon terms induced by the Kerr interaction will introduce an effective amplification into the linearized optomechanical coupling. Meanwhile, the optical bath will also be amplitied by the two-photon terms and this amplification effect can be eliminated by introducing a well designed squeezed vacuum reservoir. We have analyzed the parameter space and discussed the stability of the linearized optomechanical system. We have also studied the enhanced optomechanical entanglement. It has been found that the optomechanical entanglement can be enhanced with the coupling amplification induced by the Kerr nonlinearity. Our results will motivate the study of enhancing other optomechanical effects via quantum amplification and reservoir design.

\begin{acknowledgments}
	The authors thank Dr. Jian Huang and Dr. Deng-Gao Lai
	for helpful discussions. J.-Q.L. was supported in part by the
	National Natural Science Foundation of China (Grants No.
	12175061, No. 12575015, No. 12247105, No. 11935006, and
	No. 12421005), National Key Research and Development
	Program of China (Grant No. 2024YFE0102400), and Hunan
	Provincial Major Sci-Tech Program (Grant No. 2023ZJ1010).
	C.L. was supported in part by the Postgraduate Scientific Research Innovation Project of Hunan Province (Grant No.
	CX20250721) and the Tenglong Innovative Talent Fund of
	Hunan Normal University (Grant No. 2025TL209).
\end{acknowledgments}

\appendix*
\section{Analyses of the stability of the system}
In this appendix, we analyze the stability of the driven Kerr cavity optomechanical system described by the linearized Hamiltonian in Eq.~(\ref{Hsd}). As shown in Eq.~(\ref{Hsd}), both the effective driving detuning $\Delta_{sd}$ and the linearized optomechanical coupling strength  $G_{sd}$ depend on the steady-state displacement amplitudes $\alpha_{ss}$ and $\beta_{ss}$, which are determined by the steady-state equation
\begin{subequations}
	\begin{align}
		-\left[\frac{\kappa_a}{2}\!+\!i \Delta_c\!+\!i g_0(\beta_{ss}^*\!+\!\beta_{ss})\right] \alpha_{ss}\!-\!2 i\chi|\alpha_{ss}|^2 \alpha_{ss}\!+\!i\Omega&\!=\!0, ~\label{alphass1}\\
		-\left(i\omega _{m}\!+\!\frac{\kappa_{b}}{2}\right)\beta_{ss} \!-\!ig_{0} \vert \alpha_{ss} \vert ^{2}&\!=\!0.~\label{alphass2}
	\end{align}
\end{subequations}
From Eq.~(\ref{alphass2}), we obtain
\begin{equation}
	\beta_{ss} =-\frac{ig_{0}}{i\omega_{m}+\frac{\kappa_{b}}{2}}\vert\alpha_{ss}\vert^{2}.
\end{equation}
Substitution of $\beta_{ss}$ into Eq.~(\ref{alphass1}) yields the cubic equation ($a\neq0$)
\begin{equation}
	ay^{3}+by^{2}+cy+d=0,~\label{cubic}
\end{equation}
where $y=\vert\alpha_{ss}\vert^{2}$, and the coefficients are introduced by
\begin{align}
	a & =\frac{4g_{0}^{4}\omega_{m}^{2}}{\left(\frac{\kappa_{b}^{2}}{4}+\omega_{m}^{2}\right)^{2}}-8\chi\frac{g_{0}^{2}\omega_{m}}{\frac{\kappa_{b}^{2}}{4}+\omega_{m}^{2}}+4\chi^{2},\nonumber \\
	b & =-4\frac{\Delta_{c}g_{0}^{2}\omega_{m}}{\frac{\kappa_{b}^{2}}{4}+\omega_{m}^{2}}+4\chi\Delta_{c},\nonumber \\
	c & =\frac{\kappa_{a}^{2}}{4}+\Delta_{c}^{2},\nonumber \\
	d & =-|\Omega|^{2}.
\end{align}
The solutions of the cubic equation~(\ref{cubic}) can be expressed as~\cite{lang2012algebra}
\begin{equation}
	y_{n=1,2,3}=-\frac{b}{3a}+z^{n}\sqrt[3]{\frac{q}{2}+\sqrt{\eta}}+z^{2n}\sqrt[3]{\frac{q}{2}-\sqrt{\eta}},~\label{cubicyalpha}
\end{equation}
where $z=\exp(2i\pi/3)$, and we introduce  $\eta=(q/2)^{2}+(p/3)^{3}$ with $	p=(3ac-b^{2})/(3a^{2})$ and $q=(-2b^{3}+9abc-27a^{2}d)/(27a^{3})$.

In our analyses in the main text, the cubic equation has three solutions, but we only consider a single real root and a triple real root, which requires the coefficients to satisfy the conditions: (i) $\eta>0$; (ii) $\eta=0$ and $q=0$. In this case, the solution of the cubic equation~(\ref{cubic}) used in this work can be expressed as
\begin{equation}
	y_{3}=-\frac{b}{3a}+\sqrt[3]{\frac{q}{2}+\sqrt{\eta}}+\sqrt[3]{\frac{q}{2}-\sqrt{\eta}}.~\label{cubicy2}
\end{equation}
By substituting Eq.~(\ref{cubicy2}) into Eq.~(\ref{alphass1}), the values $\alpha_{ss}$ and $\beta_{ss}$ can be expressed as
\begin{eqnarray}
	\alpha_{ss}& =&  i\Omega\left[\frac{\kappa_{a}}{2}+i\left(\Delta_{c}-\frac{g_{0}^{2}\omega_{m}}{\omega_{m}^{2}+\frac{\kappa_{b}^{2}}{4}}y_{3}+2\chi y_{3}\right)\right]^{-1},\notag\\
	\beta_{ss} &=&-\frac{ig_{0}}{i\omega_{m}+\frac{\kappa_{b}}{2}}y_{3}.
\end{eqnarray}
Based on the above analyses as well as Eqs.~(\ref{Deltad}) and~(\ref{Deltasd}), we can obtain the dependence of both the effective driving detuing $\Delta_{sd}$ and the linearized optomechanical coupling strength $G_{sd}$ on the parameters $\chi$ and $\Delta_{c}$.

For the dynamical evolution of the linearized optomechanical system governed by the linear Langevin equation (\ref{u}), the stability condition of the system can be obtained by analyzing the eigenvalues of the coefficient matrix $\textbf{A}$ using the Routh-Hurwitz criterion~\cite{gradshteyn2014table}. For the coefficient matrix in Eq.~(\ref{eq19}), we can obtain the characteristic equation
\begin{equation}
	\det(\textbf{A}-\lambda\textbf{I})=0,
\end{equation}
which leads the characteristic polynomial
\begin{equation}
	F(\lambda) =\lambda^{4}+a_{1}\lambda^{3}+a_{2}\lambda^{2}+a_{3}\lambda+a_{4},
\end{equation}
with
\begin{eqnarray}
	a_{1} & =& \kappa_{a}+\kappa_{b},\nonumber \\
	a_{2} & =& \frac{1}{4}\kappa_{b}^{2}+\omega_{m}^{2}+\frac{1}{4}\kappa_{a}^{2}+\Delta_{sd}^{2}+\kappa_{a}\kappa_{b},\nonumber \\
	a_{3} & =& \frac{1}{4}\kappa_{a}\kappa_{b}^{2}+\kappa_{a}\omega_{m}^{2}+\frac{1}{4}\kappa_{a}^{2}\kappa_{b}+\kappa_{b}\Delta_{sd}^{2},\nonumber \\
	a_{4} & =& \left(\frac{1}{4}\kappa_{a}^{2}+\Delta_{sd}^{2}\right)\left(\frac{1}{4}\kappa_{b}^{2}+\omega_{m}^{2}\right)\notag\\
	&&-4\omega_{m}\Delta_{sd}\left[\text{Im}(G_{sd})\text{Im}(G_{sd})+\text{Re}(G_{sd})\text{Re}(G_{sd})\right].\notag\\
\end{eqnarray}	
where ``$\det(\cdot)$" denotes the determinant of a matrix. Then the Routh matrix can be expressed as
\begin{equation}
	\textbf{R}=\left(\begin{array}{cccc}
		a_{1} & 1 & 0 & 0\\
		a_{3} & a_{2} & a_{1} & 1\\
		0 & a_{4} & a_{3} & a_{2}\\
		0 & 0 & 0 & a_{4}
	\end{array}\right).
\end{equation}
According to the Routh-Hurwitz criterion, the system is stable when all the eigenvalues of the coefficient matrix $\textbf{A}$ have negative real parts. The condition requires the determinant of the Routh matrix $\textbf{R}$ to be positive, i.e., $\det(\textbf{R})>0$, which leads
\begin{align}
	a_{1}&>0,\hspace{0.2cm}a_{3}>0,\hspace{0.2cm}a_{4}>0, \notag\\ a_{1}a_{2}a_{3}  &>a_{3}^{2}+a_{1}^{2}a_{4}.
\end{align}
Thus, we can analyze the stability of the linear optomechanical system and present the steady-state parameter range in Fig.~\ref{Fig2}, based on the discriminant conditions for the single real root or triple real root solutions of the cubic equation and the Routh-Hurwitz criterion.

\nocite{*}
\bibliography{ref}

\end{document}